\documentclass[12pt]{amsart}
\usepackage{graphicx,ifthen,xspace,comment}

\numberwithin{equation}{section}
\numberwithin{figure}{section}

\newcommand{\eq}[2][]{\ifthenelse{\equal{#1}{}}{\eqNoNum{#2}}{\eqNum{\label{#1}
#2}\\}}
\newcommand{\eqs}[2][]{\ifthenelse{\equal{#1}{}}{\eqsNoNum{#2}}{\eqsNum{\label{#1}#2}\\}}
\newcommand{\eqNoNum}[1]{\begin{equation*}#1\end{equation*}}
\newcommand{\eqsNoNum}[1]{\begin{equation*}\begin{split}#1\end{split}\end{equation*}}
\newcommand{\eqNum}[2]{\begin{equation}#1\end{equation}}
\newcommand{\eqsNum}[1]{\begin{equation}\begin{split}#1\end{split}\end{equation}}
\newcommand{\xth}{\textsuperscript{th}\xspace}
\newcommand{\cc}{cost contribution\xspace}

\newcommand{\lrparen}[2][1]{\ifthenelse{\equal{#1}{1}}{(#2)}{\left(#2\right)}}
\newcommand{\N}{\mathbb{N}}
\newcommand{\Z}{\mathbb{Z}}
\newcommand{\set}[2]{\{\ifthenelse{\equal{#2}{}}{#1}{#1 : #2}\}}
\DeclareMathOperator{\subop}{sub}
\newcommand{\sub}[4][1]{\subop\lrparen[#1]{#2, #3, #4}}
\DeclareMathOperator{\detop}{det}
\renewcommand{\det}[2][1]{\detop\lrparen[#1]{#2}}
\newcommand{\smin}{\setminus}
\newcommand{\nterms}[2][1]{T\lrparen[#1]{#2}}

\begin{document}

\title[Determinants with Many Variables]{Efficient Calculation of Determinants of Symbolic Matrices with Many Variables}
\author[Tanya Khovanova]{Tanya Khovanova$^1$}\thanks{$^1$MIT}
\author[Ziv Scully]{Ziv Scully$^2$}\thanks{$^2$MIT PRIMES}
\begin{abstract}
Efficient matrix determinant calculations have been studied since the 19th century. Computers expand the range of determinants that are practically calculable to include matrices with symbolic entries. However, the fastest determinant algorithms for numerical matrices are often not the fastest for symbolic matrices with many variables. We compare the performance of two algorithms, fraction-free Gaussian elimination and minor expansion, on symbolic matrices with many variables. We show that, under a simplified theoretical model, minor expansion is faster in most situations. We then propose optimizations for minor expansion and demonstrate their effectiveness with empirical data.
\end{abstract}
\maketitle

\section{Introduction}
Determinants of square matrices are essential in a myriad of fields, both theoretical and applied. Computers can be used to quickly calculate determinants of matrices with numeric and, in more recent decades, symbolic entries. Efficient algorithms exist for matrices with a small number of variables of high degree. We turn our attention here to the opposite case: matrices of polynomials with many variables of low degree.

Bareiss introduced an algorithm, fraction-free Gaussian elimination, that requires $O(n^3)$ polynomial operations and avoids fractions \cite{bareiss}. However, the cost of those operations becomes prohibitively large. Another algorithm, minor expansion, takes $O(2^nn)$ polynomial operations and has been shown to be an attractive alternative \cite{gentleman-johnson}. Our first result, presented in Section \ref{algCompare}, is a quantitative comparison of the cost of these two algorithms on a dense matrix of linear polynomials. We find that minor expansion is favorable unless the size of the matrix is greater than some function $f$ of the number of variables $s$, which by our evidence grows faster than linearly. To the extent that our computer hardware can handle, the theoretical analysis appears reasonable when compared to experiments, though the exact correlation depends heavily on implementation details of each algorithm.

Griss introduced row-ordering as a technique to reduce further the cost of minor expansion \cite{griss1, griss2}. Our second result, presented in Section~\ref{algOptimize}, suggests the scenarios when row sorting is most helpful, namely, in our experiment, when approximately half of the entries of a matrix are 0. We try several possible sorting strategies, all of which perform similarly, so further analysis is needed to determine if any of them have an advantage over the others.

\section{Preliminaries}
We use the following notation:
\eqs{
A &= \text{an $n \times n$ matrix with entries in $\Z[x]$.} \\
a_{ij} &= \text{entry of $A$ in row $i$ and column $j$; starting index is 1.} \\
\det{A} &= \text{determinant of $A$.} \\
[a] &= \set{n \in \Z}{1 \leq n \leq a}. \\
\sub{A}{I}{J} &= \text{submatrix of $A$ using rows in $I$ and columns in $J$,} \\ &\qquad \text{where $I,J \subseteq [n]$.} \\
\nterms{p} &= \text{number of terms in polynomial $p$.}
}

Let $i \in [n]$. The determinant of $A$ can be defined recursively as
\eq[detDef]{
\det{A} =
\begin{cases}
a_{11} & \text{ if } n = 1 \\
\sum_{j=1}^{n}(-1)^{n+j}a_{ij}\det{A_{ij}} & \text{ otherwise},
\end{cases}
}
where $A_{ij} = \sub{A}{[n]\smin\{i\}}{[n]\smin\{j\}}$. Note that any choice of $i$ yields the same result, as does summing over $i$ with fixed $j$. Naively calculating $\det{A}$ requires computing $n!$ products of $n$ factors each, which would take $O((n+1)!)$ polynomial multiplications. There are several algorithms that improve drastically on this performance, two of which we study here.

\subsection{Minor expansion}\label{minExpIntro}
Calculating the determinant of $A$ requires calculating the determinant of $A_{ij} = \sub{A}{[n]\smin\{i\}}{[n]\smin\{j\}}$ for fixed $i$ and all $j \in [n]$. Similarly, the calculation of each of these determinants requires calculating the determinants of $\sub{A}{[n]\smin\{i,k\}}{[n]\smin\{j,l\}}$ for fixed distinct $i$ and $k$ and all distinct $j,l \in [n]$. Notice that, for given $j$ and $l$, the determinants of both $A_{ij}$ and $A_{il}$ require calculating the determinant of $\sub{A}{[n]\smin\{i,k\}}{[n]\smin\{j,l\}}$. Taking advantage of this and similar redundancies allows us to reduce the number of determinants we have to calculate. Increasing $i$ from 2 to $n$, we calculate determinants of every $i \times i$ submatrix contained in the first $i$ rows as a linear combination of the $(i-1) \times (i-1)$ determinants calculated in the step before. The determinants of these submatrices are called \emph{minors}. The \emph{minor expansion} algorithm follows.

\begin{quote} \bf
input $A$ \\
$M_\emptyset := 1$ \\
\textnormal{\it\% Calculate determinants of submatrices of increasing size.} \\
for $i := 2, \dots, n$ do
\begin{quote}
for $J \subseteq [n] : |J| = i$ do
\begin{quote}
\textnormal{\it\% Calculate $\det{\sub{A}{[i]}{J}}$.} \\
\textnormal{\it\% $J$ is sorted with $k$\xth element $j_k$.} \\
$M_J := \sum_{k=1}^{i}(-1)^{i+k} a_{ij_k} M_{J \smin \{j_k\}}$
\end{quote}
endfor
\end{quote}
endfor \\
\textnormal{\it\% $A$ has all $n$ columns, so $\det{A} = M_{[n]}$.} \\
return $M_{[n]}$
\end{quote}

Minor expansion requires $\sum_{i=2}^{n}(i)\binom{n}{i} = O(2^nn)$ polynomial multiplications, which, we will see, is more than fraction-free Gaussian elimination requires.

Not every entry of $A$ is involved in the same number of multiplications. Entries in the $i$\xth row are directly present in $\binom{n}{i}$ multiplications except for those in the first row, which are in $\binom{n}{2}$ multiplications. Entries in earlier rows, through their presence in minors, are indirectly involved in many more multiplications.

\subsection{Fraction-free Gaussian elimination}
It is clear from \eqref{detDef} that the determinant of a triangular matrix is the product of its diagonal entries. Ordinary Gaussian elimination, described by the iteration below, calculates a sequence of matrices $A^{(k)}$, starting with $A^{(1)} = A$, such that $A^{(n)}$ is upper triangular and $\det{A^{(k)}} = \det{A}$ for all $k \in [n]$.
\eq{
a^{(k+1)}_{ij} = a^{(k)}_{ij} - \frac{a^{(k)}_{ik}a^{(k)}_{kj}}{a^{(k)}_{kk}} \quad \forall (i,j) \in \{k+1,\dots,n\} \times \{k,\dots,n\}.
}
Unfortunately, this method requires fractions, which, especially when working with polynomials, are costly to reduce with GCD calculations and costly to let grow without reduction.

Fraction-free Gaussian elimination calculates a different sequence of matrices $A^{[k]}$, again with $A^{[1]} = A$, such that $a^{[n]}_{nn} = \det{A}$. We define $a^{[0]}_{00} = 1$. The \emph{one-step fraction-free Gaussian elimination} algorithm calculates the sequence $A^{[k]}$ with the iteration
\eq{
a^{[k+1]}_{ij} = \frac{a^{[k]}_{kk}a^{[k]}_{ij} - a^{[k]}_{ik}a^{[k]}_{kj}}{a^{[k-1]}_{k-1,k-1}} \quad \forall (i,j) \in \{k+1,\dots,n\} \times \{k,\dots,n\}.
}
It can be shown with Sylvester's Identity that the division has no remainder and that the divisor is the largest possible that guarantees this \cite{bareiss}. The algorithm requires $\sum_{i=1}^{n-1}3(n-i)^2 = O(n^3)$ polynomial multiplications and divisions.

From here on, ``Gaussian elimination'' will refer to one-step fraction-free Gaussian elimination unless otherwise specified. We note that there is a ``two-step'' variety of fraction-free Gaussian elimination, which calculates $A^{[k]}$ for only odd $k$ using a more complicated formula, and that larger step sizes are possible.

\section{Algorithm comparison}\label{algCompare}
We mentioned previously that minor expansion and Gaussian elimination take $O(2^nn)$ and $O(n^3)$ polynomial operations, respectively. However, the time taken by an individual polynomial operation can vary drastically depending on the polynomial, and minor expansion very often outperforms Gaussian elimination for a variety of reasons.

\subsection{Theoretical example}\label{theoreticalEx}
We first consider a model in which we examine a particular class of matrices and measure the cost of each algorithm in terms of integer operations. We make the following simplifying assumptions:
\begin{itemize}
\item Addition and subtraction of polynomials have zero cost.
\item Multiplication and division of polynomials $p$ and $q$ take $\nterms{p}\nterms{q}$ integer operations.
\item All integer operations have the same cost.
\end{itemize}

Choose some $s \in \N$. Suppose that every entry of $A$ is a linear polynomial in $s$ variables where each term has exactly one variable as a factor. That is, $a_{ij} = \sum_{k=1}^{s}c_{ijk}x_k$ for all $i,j \in [n]$, where all $x_k$ are variables and all $c_{ijk} \in \Z$.

The chance of having zeros as coefficients of polynomials in the following calculations is negligible, so we ignore the possibility for simplicity.

Call a polynomial \emph{$i$-homogenous} iff it has total degree $i$ in each of its terms. For example, the entries of $A$ are 1-homogenous. An $i$-homogenous polynomial has at most $\binom{i+s-1}{s-1}$ terms. Note that the product of an $i$-homogenous polynomial and a $j$-homogenous polynomial is $(i+j)$-homogenous and that the sum of two $i$-homogenous polynomials is also $i$-homogenous.

We turn our attention first to minor expansion. When calculating the $i \times i$ minors of $A$, we perform $i\binom{n}{i}$ multiplications of a 1-homogenous polynomial by an $(i-1)$-homogenous polynomial. The maximum cost of minor expansion in integer operations, denoted $C_M$, is
\eq{
C_M = s\sum_{i=2}^{n}i\binom{n}{i}\binom{i+s-2}{s-1} =
ns\sum_{i=1}^{n-1}\binom{n-1}{i}\binom{i+s-1}{s-1}.
}

The cost of Gaussian elimination can be computed similarly. When dealing with a particular entry in the $i$\xth elimination step, we do two multiplications of two $i$-homogenous polynomials and one division of a $2i$-homogenous polynomial by an $(i-1)$-homogenous polynomial. This happens to the lower $(n-i) \times (n-i)$ block of the matrix. The maximum cost of Gaussian elimination in integer operations, denoted $C_G$, is
\eq{
C_G = \sum_{i=1}^{n-1}(n-i)^2\left(2\binom{i+s-1}{s-1}^2 + \binom{2i+s-1}{s-1}\binom{i+s-2}{s-1}\right).
}

\subsection{Experimental boundary points}
\begin{figure}[tb]
\includegraphics{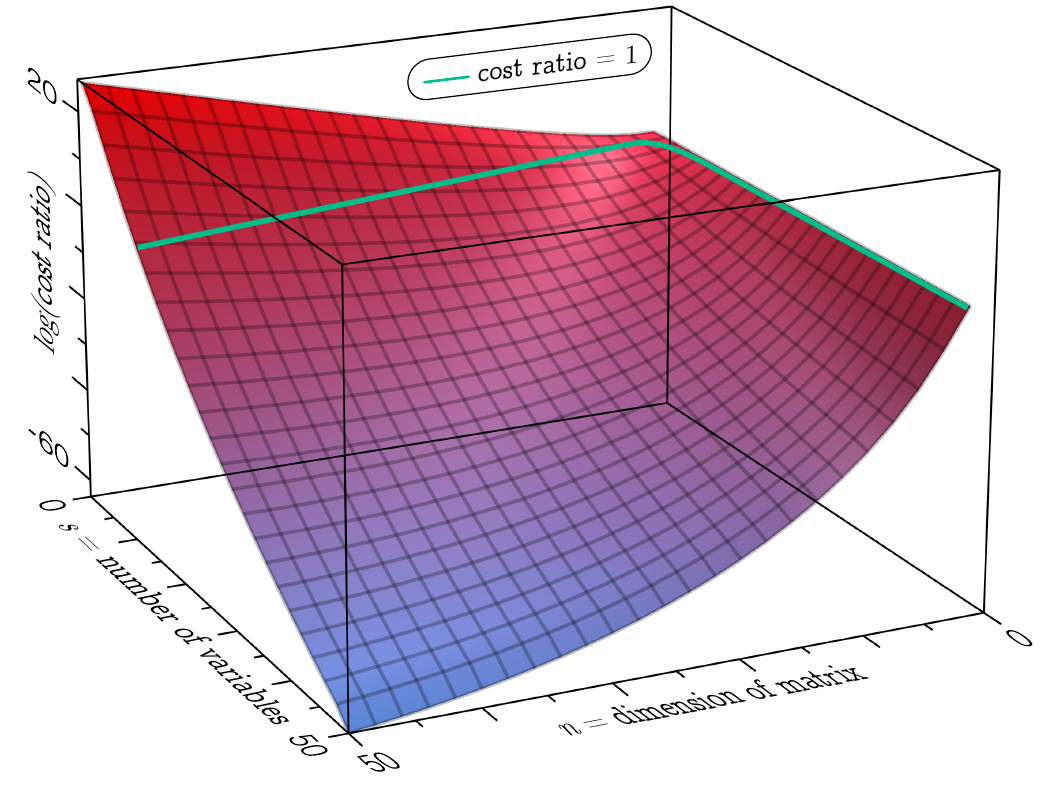}
\vspace{-.5\baselineskip}
\caption{The ratio of the cost of minor expansion to the cost of Gaussian elimination. Note that the origin is on the farthest vertical edge of the surrounding box.}
\label{costComparison}
\end{figure}

\begin{figure}[tb]
\includegraphics{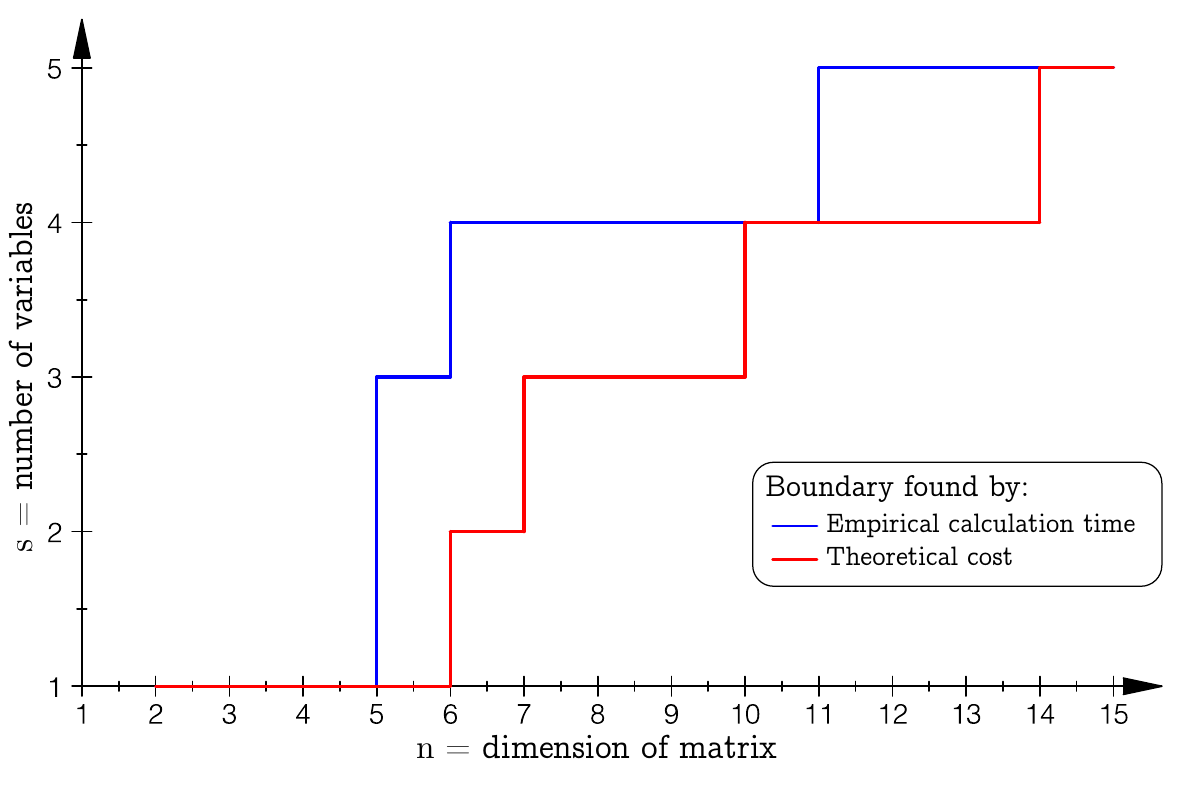}
\vspace{-.5\baselineskip}
\caption{Predicted crossover points between minor expansion and Gaussian elimination versus observed crossover points. Matrices are $n \times n$ with 1-homogenous linear polynomials in $s$ variables for entries.}
\label{theoryReality}
\end{figure}

Figure \ref{costComparison} graphs $\log(\frac{C_M}{C_G})$ with respect to $n$, the dimension of $A$, and $s$, the number of variables. Unless $n$ grows much faster than $s$, minor expansion has less costly asymptotic behavior. In the range we calculated, $n$ must grow faster than linear with respect to $s$ for Gaussian elimination to be the better choice asymptotically.

To test the predicted boundary points---points where the two algorithms have the same cost---we ran the following experiment. Starting from $n = 1$ and $s = 1$, we randomly generate $A$ and time the two determinant algorithms applied to it. When minor expansion is faster, we increment $n$; otherwise, we increment $s$. Figure \ref{theoryReality} compares the theoretically predicted boundary points with the experimentally determined ones. Unfortunately, determinants of matrices much larger than $15 \times 15$ are not easily calculable on typical home computer hardware.

Out in the wild, minor expansion tends to get the nod over Gaussian elimination when working with symbolic matrices with many variables. The analysis above suggests that minor expansion is well suited to handle many variables, and its advantage only increases when matrices are sparse. Multiplications in minor expansion always deal directly with the entries of the matrix. In contrast, the fact that Gaussian elimination involves adding terms to entries means that easy-to-multiply-by polynomials (e.g., 0, integers, monomials) not in the first row can be ``corrupted'' after an elimination step.

It is worth noting that a Gaussian elimination variant introduced by Lee and Saunders \cite{lee-saunders} takes advantage of 0s to eliminate unnecessary division. However, 0 is not the only easy-to-multiply-by polynomial. We hope that this style of cost estimation could lead to better prediction of when Gaussian elimination should be used over minor expansion, though this prediction depends on implementation details of the algorithms.

\section{Optimizing minor expansion by row permutation}\label{algOptimize}
Minor expansion is asymmetrical, passing through rows from top to bottom in the implementation given in Section \ref{minExpIntro}, which we consider in this section. We mentioned in Section \ref{algCompare} that minor expansion takes advantage of easy-to-multiply-by polynomials. Our goal in this section is to milk as much advantage out of these easy-to-multiply-by polynomials as possible.

It would be helpful to first precisely define what ``easy-to-multiply-by'' means. Making the same simplifying assumptions about multiplication as in Subsection \ref{theoreticalEx} (namely, that calculating $pq$ requires $\nterms{p}\nterms{q}$ integer operations, each of which has the same cost), we can calculate the cost of minor expansion on $A$, which we call $C_M(A)$, as
\eq{
C_M(A) = \sum_{J \subseteq [n]}\sum_{j \in J}\nterms{a_{|J|j}}\nterms{\det{\sub{A}{[|J|-1]}{J \smin \{j\}}}}.
}
Note that, although $|\det{A}|$ is invariant under row swaps and transposition of $A$, this is not in general the case for $C_M(A)$. However, in practice, calculating the cost for all $2n!$ row and column permutations would take an amount of time comparable to that of the determinant calculation itself.

It is therefore desirable to estimate the \cc of each entry or row---without taking into account other entries or rows---of a matrix and use that to sort the rows, reducing the sorting operation to $O(n^2)$ time which, in practice, is negligible compared to the determinant calculation time.

The \cc of an entry depends on:
\begin{description}
\item[Number of terms] The cost of calculating $pq$ is $\nterms{p}\nterms{q}$. If we were only performing a single multiplication, then this would be the only important attribute with regards to \cc.
\end{description}
The \cc of a row depends on:
\begin{description}
\item[A function of the number of terms of each of its entries] The sum, sum of squares, number of nonzero entries, and so on.
\item[Number of linearly independent terms] If, for example, the $i \times i$ minors all have the same form as polynomials and the $(i+1)$\xth row has $k$ linearly independent terms, then calculating the $(i+1) \times (i+1)$ minors results in, at most, $k$ terms per term of the $i \times i$ minors.
\end{description}

Recall that absolute value of the determinant of a matrix is invariant under row swaps. Given a \cc estimate for each row, sorting rows by descending \cc generally speeds up minor expansion.

To see why this is, consider three polynomials $p$, $q$ and $r$. The cost in integer multiplications of calculating $pqr$ depends on order of calculation. For instance, $(pq)r$ takes $\nterms{p}\nterms{q} + \nterms{pq}\nterms{r}$ integer multiplications. The more consolidation of like terms occurs when calculating $pq$, the further $\nterms{pq}$ is from its maximum of $\nterms{p}\nterms{q}$. A lower cost results from choosing the first two polynomials to have as much consolidation of like terms when multiplied as possible and choosing the third to have as many terms as possible. (The second goal is less obvious from these equations. It is clearly useful in the case where there is no consolidation of like terms, and it must be balanced with the first goal outside of this case.)

\begin{figure}[tb]
\includegraphics{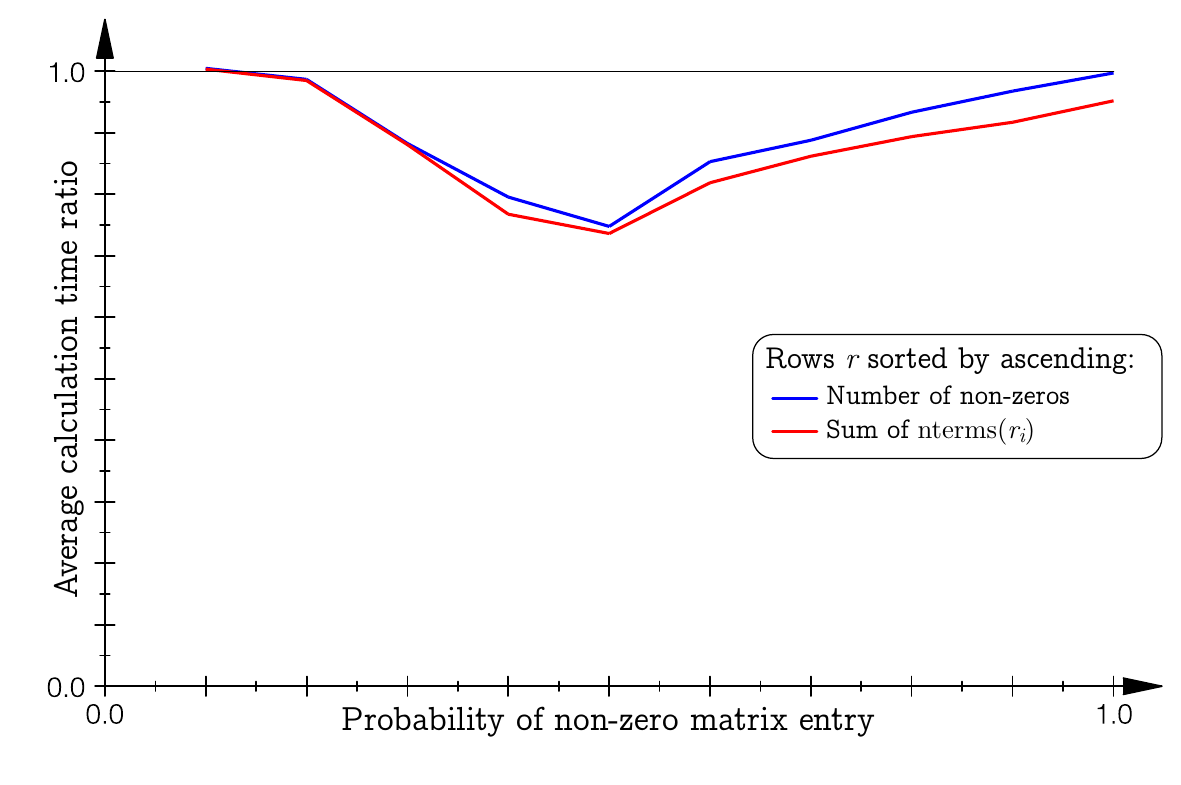}
\vspace{-\baselineskip}
\caption{The ratio of the calculation time of minor expansion after various sorting strategies to the calculation time of minor expansion without sorting. Matrices were $9 \times 9$ with random degree 1 polynomial entries in 5 variables, with integer coefficients in $[-999,999]$. Each polynomial was nonzero with probability $(1-p)$, in which case it had an equal chance of having 1, 2, 3 or 4 terms. Shows the average ratio from 100 trials of each $p$ value $0.1, 0.2, \dots, 1.0$.}
\label{sorting1}
\end{figure}

Although the \cc of rows is not yet well-defined (as, indeed, they cannot be without considering all other rows in a matrix), experiment shows that even these vague ideas are useful in practice. Our findings, shown in Figure \ref{sorting1}, are that sorting strategies are most effective when about half of a matrix's entries are 0, where sorting cut off, on average, approximately $\frac{1}{4}$ of the calculation time. We suspect that this is a reflection of the increased variance in \cc of rows.


\section*{Acknowledgements}
We thank Stefan Wehmeier of MathWorks for suggesting the problem and Ben Hinkle of MathWorks for arranging software lisences and a teleconference. We also thank MathWorks, Inc. and the MIT Program for Research In Mathematics, Engineering and Science (MIT PRIMES)

\bibliographystyle{amsplain}
\bibliography{refs.bib}

\end{document}